\documentstyle[12pt,psfig,draft,lscape,array]{nature-ejb}

\def\simlt{\mathrel{\hbox{\rlap{\hbox{\lower4pt\hbox{$\sim$}}}\hbox{$<$}}}}
\def\simgt{\mathrel{\hbox{\rlap{\hbox{\lower4pt\hbox{$\sim$}}}\hbox{$>$}}}}

\def\ale{\mathrel{\hbox{\rlap{\hbox{\lower4pt\hbox{$\sim$}}}\hbox{$<$}}}}
\def\age{\mathrel{\hbox{\rlap{\hbox{\lower4pt\hbox{$\sim$}}}\hbox{$>$}}}}

\def\ra#1#2#3{#1$^{\rm h}$#2$^{\rm m}$#3$^{\rm s}$}
\def\dec#1#2#3{$#1^\circ#2'#3''$}

\def\grb{GRB\,050724}

\newcommand{\swift}{\textit{Swift}}

\def\spose#1{\hbox to 0pt{#1\hss}}

\begin{document}

\title{\Large \bf The afterglow and elliptical host galaxy of 
the short gamma-ray burst GRB\,050724}

\author{
   E.~Berger\affiliation[1]
     {Carnegie Observatories, 813 Santa Barbara St., Pasadena, CA
        91101, USA},
   P.~A.~Price\affiliation[2]
     {Institute for Astronomy, University of Hawaii, 2680 Woodlawn
      Drive, Honolulu, HI 96822, USA},
   S.~B.~Cenko\affiliation[3]
     {Space Radiation Laboratory 220-47, California Institute of
      Technology, Pasadena, CA 91125, USA},
   A.~Gal-Yam\affiliation[4]
      {Caltech Optical Observatories 105-24, California Institute of
       Technology, Pasadena, CA 91125, USA},
   A.~M.~Soderberg\affiliationmark[4],
   M.~Kasliwal\affiliationmark[4],
   D.~C.~Leonard\affiliationmark[4],
   P.~B.~Cameron\affiliationmark[4],
   D.~A.~Frail\affiliation[5]
     {National Radio Astronomy Observatory, P.O. Box 0, Socorro, New
      Mexico 87801, USA},
   S.~R.~Kulkarni\affiliationmark[4],
   D.~C.~Murphy\affiliationmark[1],
   W.~Krzeminski\affiliation[6]
     {Las Campanas Observatory, Carnegie Observatories, Casilla 601, La
      Serena, Chile},
   T.~Piran\affiliation[7]
     {Racah Institute of Physics, Hebrew University, Jerusalem
      91904, Israel},
   B.~L.~Lee\affiliation[8]
     {Department of Astronomy and Astrophysics, University of Toronto,
      Toronto, Ontario M5S 3H8, Canada},
   K.~C.~Roth\affiliation[9]
     {Gemini Observatory, 670 N. Aohoku Place Hilo, HI 96720, USA},
   D.-S.~Moon\affiliationmark[3],
   D.~B.~Fox\affiliationmark[4],
   F.~A.~Harrison\affiliationmark[3],
   S.~E.~Persson\affiliationmark[1],
   B.~P.~Schmidt\affiliation[10]
     {RSAA, ANU, Mt.\ Stromlo Observatory, via Cotter Rd, Weston
      Creek, ACT 2611, Australia},
   B.~E.~Penprase\affiliation[11]
     {Pomona College Dept.~of Physics \& Astronomy, 610 N.~College Ave,
      Claremont, CA 91711, USA},
   J.~Rich\affiliationmark[10],
   B.~A.~Peterson\affiliationmark[10],
   and L.~L.~Cowie\affiliationmark[2]
}

\date{\today}{}
\headertitle{Afterglow and elliptical host of GRB\,050724}
\mainauthor{Berger et al.}

\summary{Despite a rich phenomenology, $\gamma$-ray bursts (GRBs) are
divided\cite{kmf+93} into two classes based on their duration and
spectral hardness -- the long-soft and the short-hard bursts.  The
discovery of afterglow emission from long GRBs was a watershed event,
pinpointing\cite{pfb+98} their origin to star forming galaxies, and
hence the death of massive stars, and indicating\cite{fks+01} an
energy release of about $10^{51}$ erg.  While theoretical
arguments\cite{kc96} suggest that short GRBs are produced in the
coalescence of binary compact objects (neutron stars or black holes),
the progenitors, energetics, and environments of these events remain
elusive despite recent\cite{bpp+05,gbb+05,gcn3585,gcn3612}
localizations.  Here we report the discovery of the first radio
afterglow from a short burst, GRB\,050724, which unambiguously
associates it with an elliptical galaxy at a redshift\cite{gcn3700},
$z=0.257$.  We show that the burst is powered by the same relativistic
fireball mechanism as long GRBs, with the ejecta possibly collimated
in jets, but that the total energy release is $10\!-\!1,000$ times
smaller.  More importantly, the nature of the host galaxy demonstrates
that short GRBs arise from an old ($>1$ Gyr) stellar population,
strengthening earlier suggestions\cite{bpp+05,gbb+05}, and providing
support for coalescing compact object binaries as the progenitors.}

\maketitle

\begin{figure} 
\centerline{\psfig{file=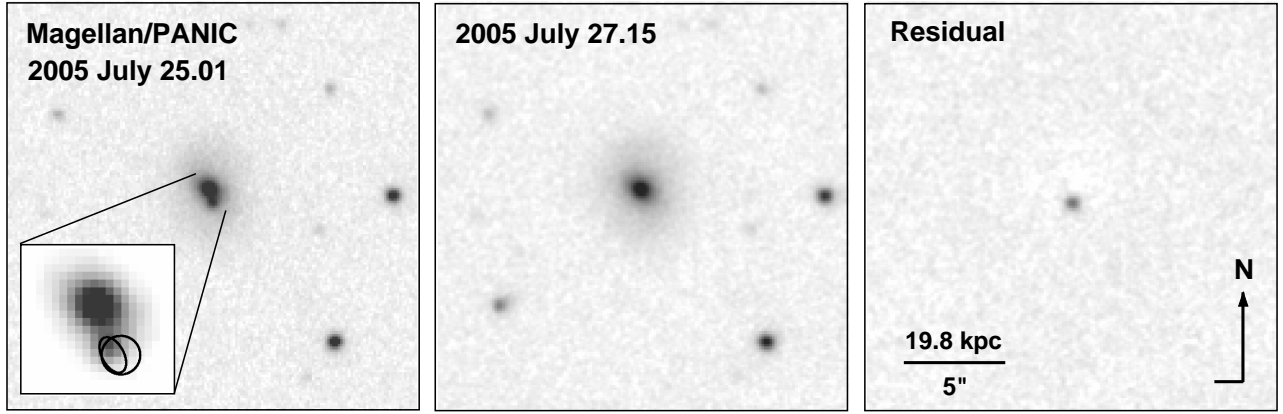,width=6.7in,angle=0}}
\caption[]{\small Near-infrared $K$-band images of the afterglow and
host galaxy of GRB\,050724.  The afterglow has completely faded
between 11.6 and 34.9 hours after the burst, indicating a flux decay
rate of $\alpha< -1.9$ ($F_\nu\propto t^\alpha$).  The position of the
NIR afterglow is $\alpha=$\ra{16}{24}{44.38},
$\delta=$\dec{-27}{32}{27.5}, with an uncertainty of about $0.1''$ in
each coordinate.  The inset shows the Very Large Array radio position
(ellipse; $\alpha=$\ra{16}{24}{44.37}, $\delta=$\dec{-27}{32}{27.5})
and the Chandra X-ray position\cite{gcn3697} (circle;
$\alpha=$\ra{16}{24}{44.36}, $\delta=$ \dec{-27}{32}{27.5}).  These
positions are fully consistent within the measurement uncertainty: the
radio-NIR offset is $\Delta\alpha=0.12\pm 0.11''$, $\Delta\delta=
0.19\pm 0.21''$, while the X-ray-NIR offset is $\Delta\alpha=0.28\pm
0.22''$ and $\Delta\delta=0.01\pm 0.22''$.  The host galaxy brightness
corrected for Galactic extinction is $K=14.88\pm 0.03$ ($3.5''$
aperture).  Using standard cosmological parameters ($\Omega_m=0.27$,
$\Omega_\lambda = 0.73$, $H_0=71$ km s$^{-1}$ Mpc$^{-1}$) the absolute
magnitude is $M_K=-24.7$, including a $k$-correction factor of $1.05$
mag.  This suggests that the host is a bright galaxy with a
luminosity, $L\approx 1.6L^*$ by comparison to the luminosity function
derived\cite{cnb+01} from 2dF and 2MASS; see also
Ref.~\pcite{gcn3700}.  The host magnitude in the optical $I$-band is
$I=18.63\pm 0.2$ mag, indicating a red $I-K\approx 2.56\pm 0.2$ mag.
The radial surface brightness distribution, measured using elliptical
isophotes, follows a de Vaucouleurs $r^{1/4}$ profile with an
effective radius of 6 kpc and a central surface brightness of
$\mu_K\approx 19.5$ mag arcsec$^{-2}$.  The ellipticity of the host is
about 0.17, indicating an E2 Hubble classification.}  
\label{fig:nir}
\end{figure}

On receipt of the \swift\ X-ray localization\cite{gcn3667} of the
short-hard burst \grb\ [duration, $T_{90}=3\pm 1$ s dominated by an
initial spike of 0.25 s; hardness ratio, $F(50\!-\!100\,{\rm
keV})/F(25\!-\!50\,{\rm kev})=0.9\pm 0.1$; Ref.~\pcite{bcb+05}] we
initiated observations in the radio, near-infrared (NIR) and optical
bands at the Very Large Array, the Baade 6.5-m Magellan telescope, and
the Swope 40-inch telescope (see Table~\ref{tab:data}).  Within the
overall uncertainty of the X-ray position we discovered a point-like
radio source and confirmed that it is the radio afterglow by its
subsequent variability.  Contemporaneous digital image subtraction of
our optical and NIR frames from the first and second nights revealed a
variable source coincident with the radio afterglow, which we identify
as the optical afterglow; this was subsequently
confirmed\cite{gcn3690} elsewhere.  The afterglow is coincident with a
bright galaxy, identified\cite{gcn3672} in earlier optical imaging,
demonstrating that it is the host galaxy; see Figure~\ref{fig:nir}.

\begin{figure}
\centerline{\psfig{file=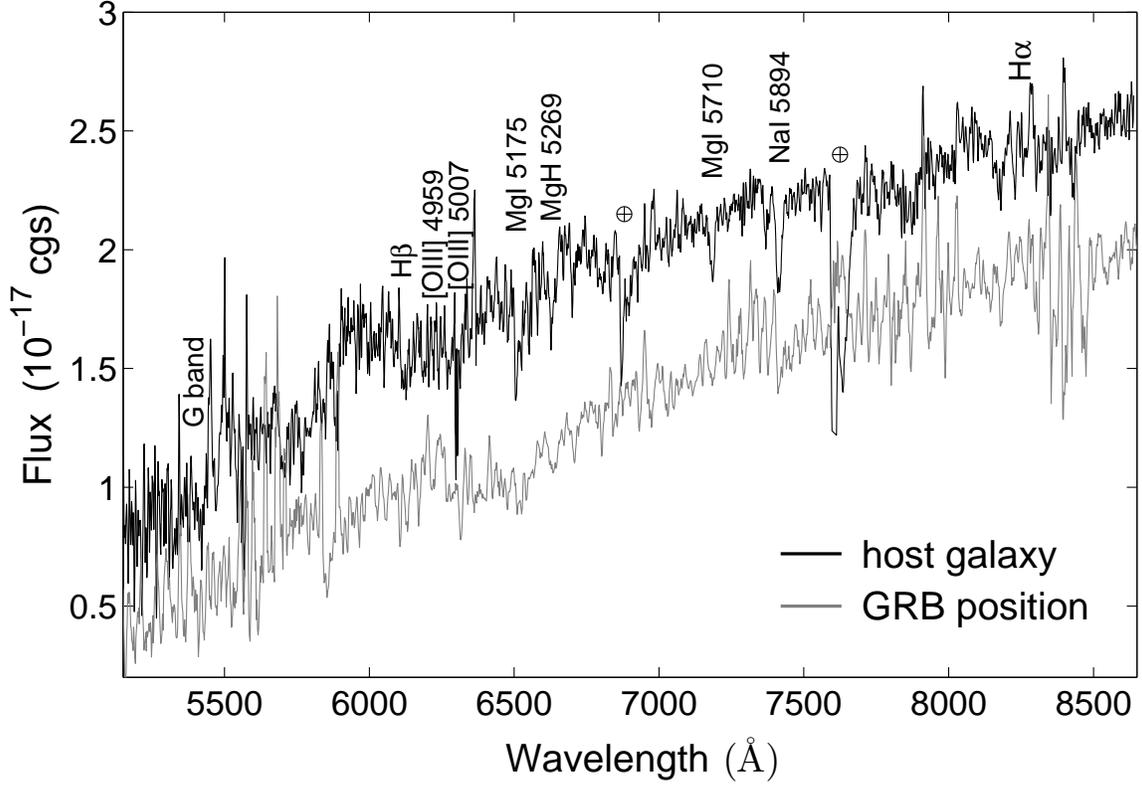,width=6in,angle=0}}
\caption[]{\small Optical spectra of the host galaxy of GRB\,050724.  
The black line shows a spectrum taken through the center of the host
galaxy with the Gemini Multi-Object Spectrograph on the Gemini North
telescope on 2005 July 27 UT.  The observations consisted of $4\times
1800$ s integrations through a $1''$ wide slit.  The spectrum was 
processed with the standard {\tt gmos} reduction tasks in IRAF before 
subtracting the sky, and extracting the spectra.  Flux calibration 
was achieved through an archival observation of a spectrophotometric 
standard star, taken with a slightly different instrumental setup; 
as such, the flux calibration is not completely accurate, but is 
indicative.  We detect the Na D lines in absorption at a mean redshift 
of $z=0.257\pm 0.001$, confirming another measurement\cite{gcn3700}.  
We also detect absorption lines corresponding to Mg\,b (5174), MgH 
(5269) and MgI (5710).  We place a $3\sigma$ limit of about $1.5\times 
10^{-17}$ erg cm$^{-2}$ s$^{-1}$ on the flux of H$\alpha$, 
corresponding at the redshift of the host to a limit of $<0.02$ 
M$_\odot$ yr$^{-1}$.  The lack of a prominent H$\beta$ absorption 
feature indicates\cite{dg83} a stellar population older than $\sim 1$ 
Gyr.  The gray line is a spectrum obtained at the position of \grb\ 
with the Low-Resolution Imaging Spectrometer on the Keck II telescope 
on 2005 July 28 UT.  A slit position angle of $90^\circ$ was chosen 
in order to minimize contribution from the galaxy light.  This allows 
us to place constraints on any residual star formation at the position 
of the GRB.  The observation consisted of a single 2240 s exposure 
with a $1.5''$ wide slit.  Flux calibration was performed using the
spectrophotometric standard BD\,+17$^\circ$4708 (red) and
BD\,+28$^\circ$4211 (blue).  The shape of the spectrum is not
completely accurate due to slit losses, but the lack of detectable 
H$\alpha$ emission allows us to place a limit of 0.03 M$_\odot$
yr$^{-1}$ on the star formation rate at the position of the GRB.}
\label{fig:spec}
\end{figure}

Motivated by this association, we used the Gemini Multi-Object
Spectrograph on the Gemini North telescope to obtain a spectrum of the
host galaxy from which we measure a redshift, $z=0.257$, confirming
other measurements\cite{gcn3700}; see Figure~\ref{fig:spec}.  At this
redshift the fluence\cite{gcn3667} of the burst, $F_\gamma\approx
6.3\times 10^{-7}$ erg cm$^{-2}$ ($15-350$ keV), translates to an
isotropic-equivalent energy release, $E_{\gamma,{\rm iso}}\approx
4\times 10^{50}$ erg; this includes a bolometric correction of a
factor of four, the average ratio of the $20\!-\!2000$ keV to the
$25\!-\!300$ keV luminosity in the BATSE short burst sample.  The
X-ray luminosity of the afterglow at $t=10$ hr, a proxy\cite{fw01} for
the kinetic energy of the blast wave, is $L_{X,{\rm iso}}\approx
4.2\times 10^{44}$ erg s$^{-1}$.  Both of these quantities are at
least an order of magnitude lower\cite{fks+01,bkf03} than for the long
GRBs.

\begin{figure}
\centerline{\psfig{file=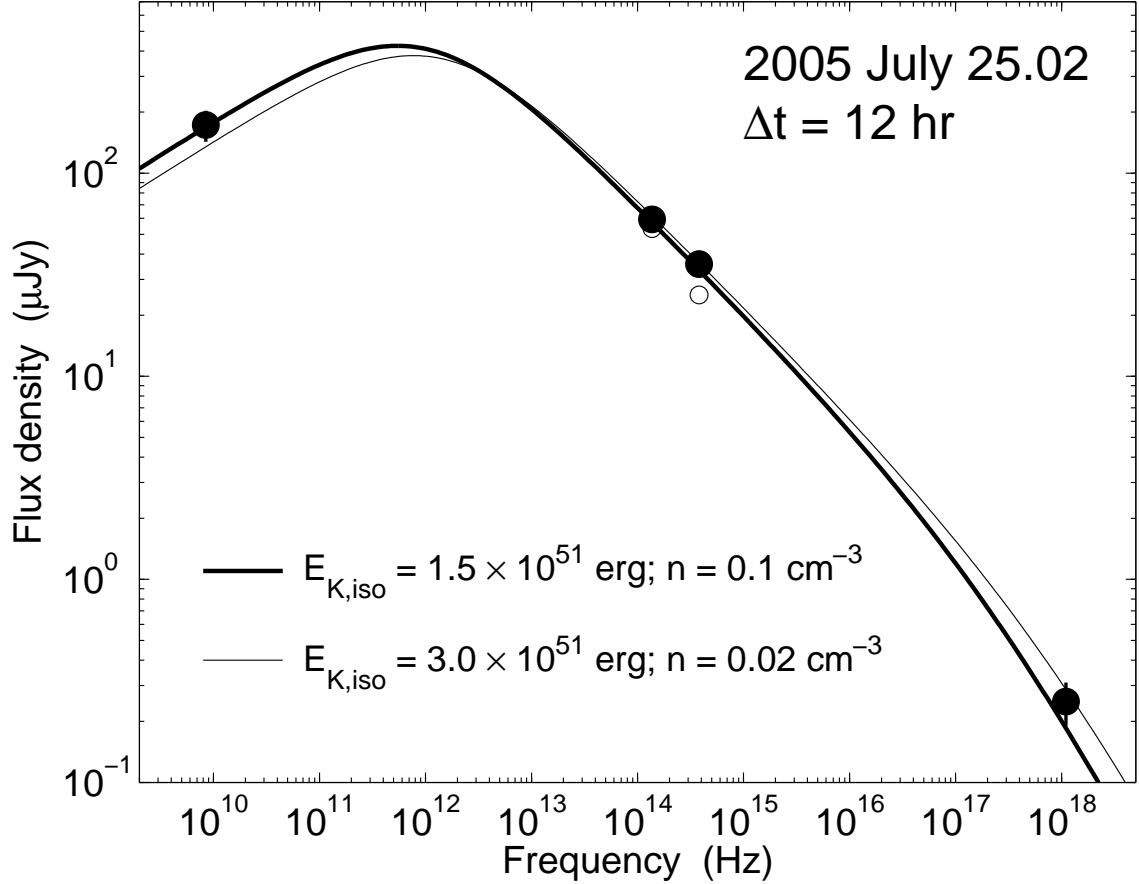,width=6in,angle=0}}
\caption[]{\small Radio to X-ray spectral energy distribution of the 
afterglow emission 12 hrs after the burst.  In the optical and NIR 
bands the open circles are the measured fluxes without a correction 
for Galactic extinction.  We find that the Galactic extinction along
the line of sight required for reconciling the optical, NIR, and 
X-ray fluxes is $A_V\approx 2.7$ mag, about $35\%$ higher than the 
tabulated value\cite{sfd98}.  The inferred extinction is in very good 
agreement with the elevated hydrogen column density, $N_H\approx 5.6
\times 10^{21}$ cm$^{-2}$, inferred\cite{bcb+05} from the X-ray 
afterglow, and it indicates that the excess absorption has a Galactic, 
rather than host galaxy, origin.  The lines are synchrotron 
models\cite{gs02} of the afterglow emission.  We find a best-fit 
solution with an energy, $E_{K,{\rm iso}}\approx 1.5\times 10^{51}$ 
erg, a density, $n\approx 0.1$ cm$^{-3}$, and fractions of energy in 
the relativistic electrons and magnetic field of $\epsilon_e\approx 
0.04$ and $\epsilon_B\approx 0.02$, respectively.  A slight degeneracy
between the energy and density is shown by the thin line, which 
marginally fits the data.}
\label{fig:bbspec}
\end{figure}

The minimum initial Lorentz factor of the ejecta is between 80 and
160, based\cite{gcn3667} on a peak flux, $f_p\approx 3.9$ ph cm$^{-2}$
s$^{-1}$, and the duration of the initial hard spike\cite{bcb+05} of
0.25 s.  This large value, similar to those inferred\cite{ls01} for
long GRBs, indicates a relativistically expanding fireball, which in
turn produces the afterglow emission.  \grb\ is the first short burst
with radio, optical/NIR, and X-ray afterglows and we are therefore in
a unique position to derive the properties of the fireball and burst
environment.  Using a standard synchrotron power-law
spectrum\cite{gs02} fit to the afterglow data at $t=12$ hr we find an
isotropic-equivalent kinetic energy, $E_{K,{\rm iso}}\approx 1.5\times
10^{51}$ erg, a density, $n\approx 0.1$ cm$^{-3}$, and fractions of
energy in the relativistic electrons and magnetic field of
$\epsilon_e\approx 0.04$ and $\epsilon_B\approx 0.02$, respectively; a
mild degeneracy between $n$ and $E_{K,{\rm iso}}$ marginally
accommodates a density as low as $0.02$ cm$^{-3}$ and an energy as
high as $3\times 10^{51}$ erg (see Figure~\ref{fig:bbspec}).  A
comparison of $E_{\gamma,{\rm iso}}$ and $E_{K,{\rm iso}}$ indicates a
radiative efficiency of about $20\%$, similar to the long GRBs.

The fading rate of the NIR afterglow emission between 12 and 35 hours
after the burst is steeper than $F_\nu\propto t^{-1.9}$
(Table~\ref{tab:data}).  This is suggestive of a collimated explosion,
or jet\cite{sph99}. The flat or rising optical light curve between 12
and 14.2 hr, suggests that the jet break time is $\sim 1$ d,
corresponding to an opening angle\cite{fks+01}, $\theta_j\sim 0.15$
rad (for $n=0.1$ cm$^{-3}$).  In this framework the true energy
release is $E_\gamma \approx 4\times 10^{48}$ erg and $E_K\approx
1.7\times 10^{49}$ erg, two orders of magnitude below the energy
release of long GRBs.  We note that jet breaks are achromatic, and our
tentative claim can be tested with additional radio data.  We conclude
from this discussion that \grb\ is powered by the same fireball
mechanism as long GRBs, with similar micro-physical properties.  The
fundamental difference is that the total energy release is a few
orders of magnitude smaller than in long GRBs.

We now turn to the nature of the progenitor system, as revealed from
the properties of the host galaxy.  The spectrum indicates that the
host is an early type galaxy\cite{gcn3700}, with a stellar population
that is older than $\sim 1$ Gyr, based on the lack of detectable
Balmer H$\beta$ absorption\cite{dg83}. From the limit on H$\alpha$
emission we find that the overall star formation rate is $<0.02$
M$_\odot$ yr$^{-1}$, and more importantly, at the location of \grb\ we
place a limit of $<0.03$ M$_\odot$ yr$^{-1}$ (Figure~\ref{fig:spec}).
The red color and surface brightness profile of the galaxy indicate an
elliptical (E2) classification.  The position of \grb\ is offset by
about $2,570\pm 80$ pc from the center of the host, corresponding to
$0.4r_e$, where $r_e\approx 6$ kpc is the host galaxy's effective
radius.  This offset is smaller\cite{bkd02} than for $80\%$ of the
long bursts.

The association of the burst with an elliptical galaxy dominated by an
old stellar population is unlike any of the long GRBs localized to
date, which invariably occur\cite{chg04} in star forming galaxies.
This leads us to conclude that the progenitor of \grb\ was not a
massive star, but was instead related to an old stellar population.
Theoretical considerations suggest\cite{kc96} that short bursts arise
from the coalescence of binary systems undergoing angular momentum
loss via gravitational radiation.  Both NS-NS and NS-BH systems have
been proposed, leading to a prediction\cite{fwh99} of a wide
distribution of coalescence timescales ($\sim 10^7-10^{10}$ yr) and
hence offsets ($\sim {\rm few}-10^3$ kpc).  For delayed mergers, the
median redshift is predicted\cite{gp05} to be lower compared to the
bulk of the star formation activity, i.e., $z\ll 1$.

The old age of the host's stellar population, the lack of detectable
current star formation at the position of the burst, and the low
redshift compared to most long GRBs point to a binary progenitor that
had a coalescence time of $\simgt 1$ Gyr.  The small offset, however,
suggests that the kick velocity imparted to the system was likely too
small to unbind it from the host.  In this context a comparison to two
recent short GRBs is illustrative.  GRB\,050509b was possibly
associated\cite{bpp+05,gbb+05} with an elliptical galaxy at $z=0.225$,
but the poor localization ($9.3''$ radius\cite{gbb+05}) also allowed
an association with higher redshift star forming galaxies.  The
results on \grb\ now lend credence to the claimed association with the
elliptical galaxy.  On the other hand, GRB\,050709 was precisely
localized within 3 kpc of a star forming galaxy\cite{gcn3605} at
$z=0.16$.  While this association does not allow a definitive argument
against a massive star origin, it suggests that the progenitors of
short GRBs occur in diverse environments, and with a range of
coalescence timescales.  This scenario is similar to that of type Ia
supernovae\cite{ber90}.

We conclude with the following intriguing possibilities.  First, the
isotropic-equivalent prompt energy release appears to correlate with
the burst duration, such that the luminosity is nearly constant,
$L_{\rm \gamma,iso}\approx (3-15)\times 10^{50}$ erg.  This may
explain why the afterglow of the 40-ms duration GRB\,050509b was
significantly fainter than those of GRBs 050709 and 050724.  Second,
the small offsets and low redshifts of the latter two bursts may
contrast with population synthesis models, which
predict\cite{fwh99,gp05} $90\%$ of the offsets to be $>10$ kpc, and a
median redshift $z\sim 0.5-1$.  Finally, if the beaming inferred for
\grb\ is typical of the short burst population, then this implies that
the true event rate of short bursts is about fifty times higher than
observed.  This suggests that $\simlt 10\%$ of NS-NS and NS-BH
binaries\cite{phi91} end their lives in GRB explosions.  The continued
detection of short GRBs by \swift, will allow these conclusions to be
tested through the distribution of energies, jet angles, and offsets.


\begin{thebibliography}{10}
 
\bibitem[{Kouveliotou} {\it et~al.}<1>]{kmf+93}
{Kouveliotou}, C., {Meegan}, C.~A., {Fishman}, G.~J., {Bhat}, N.~P., {Briggs},
  M.~S. {\it et al.} {Identification of two classes of gamma-ray bursts}.
\newblock {\it Astrophys. J.} {\bf 413}, L101--L104, (1993).
 
\bibitem[{Pian} {\it et~al.}<2>]{pfb+98}
{Pian}, E., {Fruchter}, A.~S., {Bergeron}, L.~E., {Thorsett}, S.~E.,
  {Frontera}, F. {\it et al.} {Hubble Space Telescope Imaging of the Optical
  Transient Associated with GRB 970508}.
\newblock {\it Astrophys. J.} {\bf 492}, L103--L106, (1998).
 
\bibitem[{Frail} {\it et~al.}<3>]{fks+01}
{Frail}, D.~A., {Kulkarni}, S.~R., {Sari}, R., {Djorgovski}, S.~G., {Bloom},
  J.~S. {\it et al.} {Beaming in Gamma-Ray Bursts: Evidence for a Standard
  Energy Reservoir}.
\newblock {\it Astrophys. J.} {\bf 562}, L55--L58, (2001).
 
\bibitem[{Katz} \& {Canel}<4>]{kc96}
{Katz}, J.~I. \& {Canel}, L.~M. {The Long and the Short of Gamma-Ray Bursts}.
\newblock {\it Astrophys. J.} {\bf 471}, 915--920, (1996).
 
\bibitem[{Bloom} {\it et~al.}<5>]{bpp+05}
{Bloom}, J.~S., {Prochaska}, J.~X., {Pooley}, D., {Blake}, C.~W., {Foley},
  R.~J. {\it et al.} {Closing in on a Short-Hard Burst Progenitor: Constraints
  from Early-Time Optical Imaging and Spectroscopy of a Possible Host Galaxy of
  GRB 050509b}.
\newblock {\it astro-ph/0505480}, (2005).
 
\bibitem[{Gehrels}<6>]{gbb+05}
{Gehrels}, N. {The first localization of a short gamma-ray burst by Swift}.
\newblock {\it astro-ph/0505107}, (2005).
 
\bibitem[Fox {\it et~al.}<7>]{gcn3585}
Fox, D.~B., Frail, D.~A., Cameron, P.~B.  \& Cenko, S.~B. 
{GRB050709: Candidate X-ray Afterglow from Chandra}.
\newblock {\it GRB Circular Network} {\bf 3585}, (2005).
 
\bibitem[Price {\it et~al.}<8>]{gcn3612}
Price, P.~A., Jensen, B.~L., Jargensen, U.~G., Hjorth, J., Fynbo, P.~U. {\it et
  al.} {GRB 050709: optical afterglow candidate}.
\newblock {\it GRB Circular Network} {\bf 3612}, (2005).
 
\bibitem[Prochaska {\it et~al.}<9>]{gcn3700}
Prochaska, J.~X., Bloom, J.~S., Chen, H.-W., Hansen, B., Kalirai, J. {\it et
  al.} {GRB 050724: Secure Host Redshift from Keck}.
\newblock {\it GRB Circular Network} {\bf 3700}, (2005).
 
\bibitem[Krimm {\it et~al.}<10>]{gcn3667}
Krimm, H., Barbier, L., Barthelmy, S., Cummings, J., Fenimore, E. {\it et al.}
  {GRB050724: Refined analysis of the Swift-BAT possible short burst}.
\newblock {\it GRB Circular Network} {\bf 3667}, (2005).
 
\bibitem[Barthelmy {\it et~al.}<11>]{bcb+05}
Barthelmy, S.~D., Chincaini, G., Burrows, D.~N., Gehrels, N., Covino, S. 
{\it et al.} {Unravelling the origin of short gamma-ray bursts}.
\newblock {\it Submitted to Nature} (2005).
 
\bibitem[D'Avanzo {\it et~al.}<12>]{gcn3690}
D'Avanzo, P., Covino, S., Antonelli, L.~A., Melandri, A., Malesani, D. {\it et
  al.} {GRB050724: VLT observations of the variable source}.
\newblock {\it GRB Circular Network} {\bf 3690}, (2005).
 
\bibitem[Bloom {\it et~al.}<13>]{gcn3672}
Bloom, J.~S., Dupree, A., Chen, H.-W.  \& Prochaska, J.~X. {GRB050724: GMOS
  Imaging and Spectroscopy}.
\newblock {\it GRB Circular Network} {\bf 3679}, (2005).
 
\bibitem[{Freedman} \& {Waxman}<14>]{fw01}
{Freedman}, D.~L. \& {Waxman}, E. {On the Energy of Gamma-Ray Bursts}.
\newblock {\it Astrophys. J.} {\bf 547}, 922--928, (2001).
 
\bibitem[{Berger}, {Kulkarni} \& {Frail}<15>]{bkf03}
{Berger}, E., {Kulkarni}, S.~R.  \& {Frail}, D.~A. {A Standard Kinetic Energy
  Reservoir in Gamma-Ray Burst Afterglows}.
\newblock {\it Astrophys. J.} {\bf 590}, 379--385, (2003).
 
\bibitem[{Lithwick} \& {Sari}<16>]{ls01}
{Lithwick}, Y. \& {Sari}, R. {Lower Limits on Lorentz Factors in Gamma-Ray
  Bursts}.
\newblock {\it Astrophys. J.} {\bf 555}, 540--545, (2001).
 
\bibitem[{Granot} \& {Sari}<17>]{gs02}
{Granot}, J. \& {Sari}, R. {The Shape of Spectral Breaks in Gamma-Ray Burst
  Afterglows}.
\newblock {\it apj} {\bf 568}, 820--829, (2002).
 
\bibitem[{Sari}, {Piran} \& {Halpern}<18>]{sph99}
{Sari}, R., {Piran}, T.  \& {Halpern}, J.~P. {Jets in Gamma-Ray Bursts}.
\newblock {\it Astrophys. J.} {\bf 519}, L17--L20, (1999).
                                                         
\bibitem[{Dressler} \& {Gunn}<19>]{dg83}
{Dressler}, A. \& {Gunn}, J.~E. {Spectroscopy of galaxies in distant clusters.
  II - The population of the 3C 295 cluster}.
\newblock {\it Astrophys. J.} {\bf 270}, 7--19, (1983).
 
\bibitem[{Bloom}, {Kulkarni} \& {Djorgovski}<20>]{bkd02}
{Bloom}, J.~S., {Kulkarni}, S.~R.  \& {Djorgovski}, S.~G. {The Observed Offset
  Distribution of Gamma-Ray Bursts from Their Host Galaxies: A Robust Clue to
  the Nature of the Progenitors}.
\newblock {\it Astron. J.} {\bf 123}, 1111--1148, (2002).
 
\bibitem[{Christensen}, {Hjorth} \& {Gorosabel}<21>]{chg04}
{Christensen}, L., {Hjorth}, J.  \& {Gorosabel}, J. {UV star-formation rates of
  GRB host galaxies}.
\newblock {\it Astr. Astrophys.} {\bf 425}, 913--926, (2004).
 
\bibitem[{Fryer}, {Woosley} \& {Hartmann}<22>]{fwh99}
{Fryer}, C.~L., {Woosley}, S.~E.  \& {Hartmann}, D.~H. {Formation Rates of
  Black Hole Accretion Disk Gamma-Ray Bursts}.
\newblock {\it Astrophys. J.} {\bf 526}, 152--177, (1999).
 
\bibitem[{Guetta} \& {Piran}<23>]{gp05}
{Guetta}, D. \& {Piran}, T. {The luminosity and redshift distributions of
  short-duration GRBs}.
\newblock {\it Astr. Astrophys.} {\bf 435}, 421--426, (2005).
 
\bibitem[Price<24>]{gcn3605}
Price, P.~A. {GRB 050709: Spectroscopy}.
\newblock {\it GRB Circular Network} {\bf 3605}, (2005).
 
\bibitem[{van den Bergh}<25>]{ber90}
{van den Bergh}, S. {The frequency of SN IA in galaxies of different Hubble
  type}.
\newblock {\it Publ. Astr. Soc. Pacific} {\bf 102}, 1318--1320, (1990).
 
\bibitem[{Phinney}<26>]{phi91}
{Phinney}, E.~S. {The rate of neutron star binary mergers in the universe -
  Minimal predictions for gravity wave detectors}.
\newblock {\it Astrophys. J.} {\bf 380}, L17--L21, (1991).
 
\bibitem[{Barris} {\it et~al.}<27>]{btn+05}
 {Barris}, B.~J., {Tonry}, J.~L., {Novicki}, M.~C.  \& {Wood-Vasey}, W.~M. {The
 NN2 Flux Difference Method for Constructing Variable Object Light Curves}.
\newblock {\it astro-ph/0507584}, (2005).
 
\bibitem[{Schlegel}, {Finkbeiner} \& {Davis}<28>]{sfd98}
{Schlegel}, D.~J., {Finkbeiner}, D.~P.  \& {Davis}, M. {Maps of Dust Infrared
  Emission for Use in Estimation of Reddening and Cosmic Microwave Background
  Radiation Foregrounds}.
\newblock {\it Astrophys. J.} {\bf 500}, 525--553, (1998).
 
\bibitem[Burrows {\it et~al.}<29>]{gcn3697}
Burrows, D., Grupe, G., Kouveliotou, C., Patel, S., Meszaros, P. {\it et al.}
  {GRB 050724: Chandra Observations of the X-ray Afterglow}.
\newblock {\it GRB Circular Network} {\bf 3697}, (2005).
 
 
\bibitem[{Cole} {\it et~al.}<30>]{cnb+01}
{Cole}, S., {Norberg}, P., {Baugh}, C.~M., {Frenk}, C.~S., {Bland-Hawthorn}, J.
  {\it et al.} {The 2dF galaxy redshift survey: near-infrared galaxy luminosity
  functions}.
\newblock {\it Mon. Not. R. astr. Soc.} {\bf 326}, 255--273, (2001).
 
\end{thebibliography}

\begin{acknowledge} 
We are, as always, indebted to Scott Barthelmy and the GCN.  GRB
research at Carnegie and Caltech is supported in part by funds from
NASA.  E.B.~and A.G.~are supported by NASA through Hubble Fellowship
grants awarded by the Space Telescope Science Institute, which is
operated by AURA, Inc., for NASA.  The VLA is operated by the National
Radio Astronomy Observatory, a facility of the National Science
Foundation operated under cooperative agreement by Associated
Universities, Inc.
\end{acknowledge}

\noindent{\bf Author information} The authors declare no competing 
financial interests.  Correspondence and requests for materials
should be addressed to E.B.~(eberger@ociw.edu).

\clearpage
 
\begin{table}
\begin{center}
\begin{tabular}{>{\scriptsize}l >{\scriptsize}c >{\scriptsize}l
>{\scriptsize}l >{\scriptsize}l}
\hline
\hline
\small Epoch & $\Delta t$ & Telescope & Band & Flux \\
       (UT)  & (hr)       &           &      & ($\mu$Jy) \\\hline
Jul 25.01 & 11.6 & Magellan/PANIC & $K$ & $38.7\pm 1.4$ \\
Jul 25.98 & 34.9 & Magellan/PANIC & $K$ & $<4.6$ \\
Jul 25.02 & 12.0 & Swope 40-inch  & $I$ & $8.4\pm 0.2$ \\
Jul 25.11 & 14.2 & Swope 40-inch  & $I$ & $11.1\pm 0.9$ \\
Jul 26.05 & 36.7 & Swope 40-inch  & $I$ & $<4.0$ \\
Jul 25.09 & 13.7 & VLA & 8.46 & $173\pm 30$ \\
Jul 26.21 & 40.5 & VLA & 8.46 & $465\pm 29$ \\\hline
\end{tabular}
\end{center}
\caption[]{\small Afterglow observations of GRB\,050724 in the radio,
optical, and near-infrared bands.  For the radio observations we list
the observing band in GHz, while for the optical and NIR data we list
the filter.  In all VLA observations we used the extra-galactic 
sources 3C\,286 and J1626-298 for flux and phase calibration, 
respectively.  The data were reduced and analyzed using the 
Astronomical Image Processing System, and the flux density and 
uncertainty were measured from the resulting maps.  The NIR data were 
obtained with Persson's Auxilliary Nasmyth Infrared Camera and 
consisted of sixty-six 20-s images at each epoch.  The individual 
images were processed in the standard manner and corrected for 
distortion.  Astrometry was performed relative to nine 2MASS sources 
resulting in an astrometric accuracy of about $0.1''$.  The optical 
observations consisted of two 900-s images in the first two epochs 
and three 900-s images in the third epoch.  The data were processed 
in the standard manner, and astrometry was performed relative to 65 
USNO stars, resulting in an rms uncertainty of about $0.15''$.  
Photometry of the NIR afterglow was performed using the ``NN2'' 
method\cite{btn+05}.  Errors in the subtracted fluxes were measured 
from the rms deviation of fluxes within apertures randomly distributed 
over the background of the subtracted images.  Flux calibration was 
performed relative to sources from the 2MASS catalog; our resulting 
absolute calibration is limited by statistical errors in the faint 
catalog sources to an accuracy of about $5\%$.  Photometry of the 
optical afterglow was performed using image subtraction, and the 
flux measurements carry a systematic uncertainty of about $0.15$ mag, 
which affects all epochs in the same manner.  Therefore, the observed 
re-brightening between 12 and 14.2 hr is significant at about 
$3\sigma$.  The optical and NIR fluxes are not corrected for Galactic 
extinction (see Figure~\ref{fig:bbspec}).}
\label{tab:data}
\end{table}

\clearpage

\clearpage

\clearpage

\end{document}